# Mechanism-Guided Residual Lifting and Control Consistent Modeling for Pneumatic Drying Processes

Yue WU *

Xinjiang Cigarette Factory Hongyun Honghe Tobacco (Group) Co., Ltd,Urumqi, 830000, China

School of Automation, Xi'an Jiaotong University, Xi'an, 710049, China

wuyue0619@xjtu.edu.cn

*Abstract*—Pneumatic drying processes in industries such as agriculture, chemicals, and pharmaceuticals are notoriously difficult to model and control due to multi-source disturbances, coupled stage dynamics, and significant measurement delays. Traditional modeling paradigms often fail to simultaneously deliver accuracy, interpretability, and closed-loop applicability. To address this challenge, this paper introduces a unified hybrid modeling framework, termed Physics-Guided Residual Lifting with Control-Consistent Correction (PGRL-CC), which integrates a transient mechanistic model with a stability-constrained data-driven component. The framework covers the complete process chain of drying, transport, and winnowing. On the mechanistic level, the model unifies mass transfer dynamics using the partial pressure difference of water vapor, incorporates water activity clamping and latent heat corrections for bound water, and ensures energy closure with moisture-dependent specific heat. On the data-driven level, we propose an orthogonal residual learning scheme. It leverages intermediate states from the mechanistic model as proxy variables to construct a physics-inspired dictionary, preventing parameter compensation and overfitting during ridge regression. Furthermore, to ensure suitability for predictive control, a Control-Consistent Extended Dynamic Mode Decomposition with stability constraints (CC-EDMDc-S) is employed to learn the residual dynamics, for which we provide boundedness proofs and stability guarantees. The framework was validated on 10 industrial batches, comprising 63,000 samples. On unseen test data, the hybrid model achieved a Mean Absolute Error (MAE) of 0.016% for outlet moisture and 0.015 °C for outlet temperature, with $R^2$ values improving to 0.986 and 0.995, respectively. The resulting prediction residuals exhibit white-noise characteristics, with significantly reduced spectral energy at low frequencies. This research demonstrates a methodology that preserves mechanistic interpretability while substantially enhancing generalization and industrial utility, providing a replicable foundation for online optimization, quality control, and the development of digital twins.

## I. Introduction

Pneumatic drying is a highly efficient unit operation widely employed for processing granular materials, filter cakes, and slurries in agricultural, chemical, and pharmaceutical industries. Its exceptional heat and mass transfer efficiency and short material residence times are pivotal for enhancing productivity while maintaining product quality. However, as noted by Martynenko et al., drying processes universally face significant challenges, including strong nonlinearities, unknown dynamics, and the difficulty of real-time measurement of key parameters. These issues are amplified in pneumatic drying, leading to common industrial pain points such as large fluctuations in product quality and low energy efficiency.

These challenges are rooted in the inherently complex dynamics of the process. The system is characterized by strong coupling and nonlinearity arising from gas-solid two-phase flow, turbulent heat and mass transfer, and phase transitions. The process is continually subjected to multi-source disturbances, such as fluctuations in feed moisture, feed rate, and hot air conditions. Finally, a fundamental mismatch exists between the process timescale and traditional control strategies. Materials reside in the dryer for only a few seconds, whereas conventional feedback control, reliant on outlet measurements, suffers from significant pure time delay, rendering control actions often too late. This intrinsic conflict necessitates that any effective control strategy must possess predictive capabilities, creating an urgent demand for advanced process control.

To achieve effective prediction and control, two primary modeling avenues have been explored: mechanistic (first-principles) modeling and data-driven modeling. Mechanistic models, while offering deep physical insight, often suffer from inaccuracy due to oversimplification or become computationally prohibitive for real-time control. Data-driven models can fit specific operating conditions well but heavily depend on vast amounts of data, lack physical interpretability, and exhibit poor extrapolation capabilities, limiting their performance when integrated with optimization-based controllers exploring novel operating regimes.

To overcome the limitations of these individual paradigms, hybrid modeling has emerged as a promising direction. However, a simple superposition of physics and data does not resolve the fundamental issues. Existing hybrid methods still face challenges, including the dependency of residuals on unmeasurable states, unclear division of responsibilities between the physical and data components, and a disconnect between model accuracy and closed-loop control performance.

To address these difficulties, this paper introduces a novel framework named Physics-Guided Residual Lifting & Control-Consistent Extended Dynamic Mode Decomposition (**PGRL-CC**). This framework is specifically designed to construct a high-fidelity, interpretable dynamic model tailored for industrial control applications. It systematically integrates a transient mechanistic model with a data-driven component that learns the residual dynamics in a lifted feature space, guided by Koopman operator theory.

The core contributions of this work are threefold:

**Physics-Guided Residual Lifting (PGRL):** To address the issue of unmeasurable intermediate variables, we propose a proxy-state residual modeling approach. This method utilizes the intermediate and final predicted states generated by the mechanistic model itself as information-rich proxies for the true internal states of the system. This allows the data-driven component to learn and predict the residual patterns of the mechanistic model across different operating regimes.

**Orthogonal Learning and Control-Consistent Weighting:** We introduce an orthogonal projection technique that cleanly separates the learning tasks of the mechanistic model (capturing slow, global trends) and the residual model (capturing fast, local dynamics). This prevents the data-driven component from compensating for identifiable physical phenomena. Furthermore, we incorporate a control-consistent weighting scheme in the training objective, prioritizing accuracy in regions critical to closed-loop performance, such as near operational constraints.

**Stability-Constrained Dynamic Modeling (EDMDc-S):** To ensure the model is suitable for multi-step prediction in controllers like Model Predictive Control (MPC), we employ an Extended Dynamic Mode Decomposition with control and stability constraints (EDMDc-S). This method learns a linear predictor for the residual dynamics in a lifted state-space while explicitly enforcing contractiveness, guaranteeing bounded and stable long-horizon rollouts.

## II. Transient Mechanistic Model of Pneumatic Conveying Drying

To construct a mechanistic model capable of accurately describing the transient dynamics of the pneumatic conveying drying process, the system is decomposed into three sequential sections with distinct physical characteristics, as depicted in Figure 1.

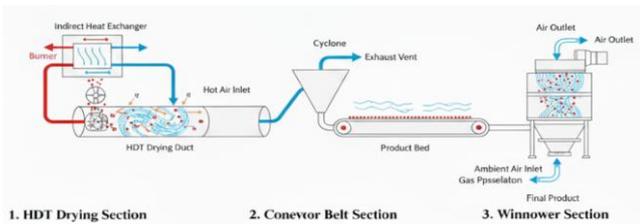

**Figure 1: the Pneumatic Drying Tube (HDT) Section, the Conveyor Belt Section, and the Winnower Section.**

### A. Material Properties and Equilibrium Relationships

The evaporation process is driven by the difference in water vapor partial pressure between the material surface and the surrounding air. The upper limit of this driving force is determined by the material temperature $T$, with the saturation vapor pressure of pure water, $P_{sat}$, calculated using the Magnus formula (with temperature in Kelvin):

$$P_{sat}(T) = 610.78 \exp\left(\frac{17.27(T - 273.15)}{237.7 + (T - 273.15)}\right) \quad [Pa] \quad (1)$$

However, the water within the material is not entirely free water. Water activity ($a_w < 1$) describes the degree to which bound water is constrained, effectively reducing the actual water vapor pressure at the material surface. This parameter is critical for accurately modeling the later stages of drying:

$$a_w(T, X) = f(X) \exp\left[\theta_{aw1}\left(\frac{1}{T} - \frac{1}{298.15}\right)\right],$$

$$a_w^* = \min\{1, \max\{0, a_w\}\} \quad (2)$$

where $f(X)$ is a material-specific sorption isotherm model. The clamped value $a_w^*$ is used in computations to ensure physical bounds.

The temperature change of the material depends on the balance of heat input and output. The specific heat of the wet material, $C_p$, which defines its thermal inertia, is calculated based on a mass-weighted average:

$$C_p(X) = \frac{\theta_{Cp1} + X \cdot \theta_{Cp2}}{1 + X} \quad (3)$$

Evaporation is the primary energy-consuming term. The latent heat of vaporization, $L_v$, not only varies with temperature but also increases in the presence of bound water to account for the additional heat of sorption required to overcome intermolecular forces:

$$L_v(T, a_w) = [2501 - 2.361(T - 273.15)] \times 10^3 \cdot [1 + \theta_{Lv,aw}(1 - a_w)] \quad (4)$$

The endpoint of the drying process is reached when the material's moisture content equilibrates with the ambient temperature and humidity. This equilibrium moisture content, $X_e$, is described by the modified Oswin model:

$$X_e(T, RH) = (\theta_{xe1} + \theta_{xe2} T) \left(\frac{RH}{1 - RH}\right)^{\theta_{xe3}} \quad (5)$$

In the later drying stages, the rate of moisture migration from the material's interior to its surface becomes the bottleneck. This rate is governed by the effective diffusion coefficient, $D_{eff}$, which incorporates the combined effects of temperature (Arrhenius dependency), moisture content, and external relative humidity:

$$D_{eff}(T, RH, X) = \theta_{D0} \exp\left(-\frac{\theta_{Ea}}{R_{gas} T}\right)(1 + 2X)[1 - \theta_{D,RH} RH] \quad (6)$$

### B. Pneumatic Drying Tube (HDT) Section

In this section, the material is conveyed by high-velocity hot air, resulting in extremely high heat and mass transfer efficiencies. The drying is dominated by the evaporation of surface moisture.

**Equivalent Area:** Assuming cylindrical particles, the surface-area-to-volume ratio is used to calculate the total heat transfer area $A_{HDT}(t)$:

$$\frac{S}{V}\Big|_{\text{cylinder}} = \frac{2}{r} + \frac{2}{L_t}, \quad A_{HDT}(t) = \frac{M_{wet}(t)}{\rho_{mat}}\left(\frac{2}{r} + \frac{2}{L_t}\right) \quad (7)$$

**Heat Transfer:** Convective heat transfer from the hot air to the material is given by:

$$\dot{Q}_{conv} = h_{HDT} A_{HDT} (T_{\text{hot}} - T) \quad (8)$$

**Mass Transfer and Evaporation Rate:** The mass transfer is driven by the difference in vapor density between the material surface and the bulk gas phase. The total pressure of the drying air is denoted by $P_{total}$.

$$\rho_{v,\text{surf}} = \frac{a_w P_{sat}(T)}{R_{water} T}, \quad P_{v,\text{gas}} = \frac{w P_{total}}{0.622 + w}, \quad \rho_{v,\text{gas}} = \frac{P_{v,\text{gas}}}{R_{water} T_{\text{hot}}} \quad (9)$$

The evaporation rate is then:

$$\frac{dm_{evap}}{dt} = k_{HDT} A_{HDT} \max(0, \rho_{v,\text{surf}} - \rho_{v,\text{gas}}) \quad (10)$$

**Coupled Mass and Energy Balance:** The rate of change of the material's dry-basis moisture content, $X$, is a direct mass balance:

$$\frac{dX}{dt} = -\frac{1}{M_{dry}} \frac{dm_{evap}}{dt} \quad (11)$$

The transient energy balance couples heat input from convection with heat loss due to evaporation (both latent and sensible heat):

$$(M_{dry}C_{p,dry} + M_{water}C_{p,w}) \frac{dT}{dt}$$
$$= h_{HDT} A_{HDT}(T_{\text{hot}} - T)$$
$$- \left(L_v + C_{p,w}(T - T_{ref})\right) \frac{dm_{evap}}{dt} \quad (12)$$

### C. Conveyor Belt Section

After exiting the pneumatic dryer, the material is deposited onto a conveyor belt in a layer. External convection is reduced, and internal moisture diffusion becomes the drying bottleneck. The material also cools via natural convection with the ambient air.

**Diffusion-Dominated Drying：**

$$\frac{dX}{dt} = -\chi_{layer} D_{eff}(T, RH, X) \frac{X - X_e(T_{air}, RH)}{L_{char}^2} \quad (13)$$

Here, $\chi_{layer}$ is a geometric factor related to the packing of the material layer, and $L_{char}$ is the characteristic diffusion length.

**Heat Transfer and Energy Closure:**

$$(M_{dry}C_{p,dry} + M_{water}C_{p,w}) \frac{dT}{dt}$$
$$= -h_{conveyor} A_{\exp}(T - T_{air})$$
$$- \left(L_v + C_{p,w}(T - T_{ref})\right) M_{dry} \left(-\frac{dX}{dt}\right) \quad (14)$$

The energy balance is similar to the HDT section, but the heat source is now natural convection cooling to the ambient air, and the evaporation rate ($dX/dt$) is determined by internal diffusion.

### D. Winnower Section

Finally, materials from different drying paths are mixed in the winnower section, where rapid surface flash evaporation occurs under negative pressure or turbulent airflow.

**Inlet Mixing:**

$$X_{in} = \beta_{top} X_{conv} + (1 - \beta_{top}) X_{HDT} \quad (15)$$

$$T_{in} = \frac{\beta_{top} C_p(X_{conv}) T_{conv} + (1 - \beta_{top}) C_p(X_{HDT}) T_{HDT}}{\beta_{top} C_p(X_{conv}) + (1 - \beta_{top}) C_p(X_{HDT})} \quad (16)$$

where $\beta_{top}$ represents the mass fraction of the top layer of material from the conveyor.

**Flash Evaporation Model:** A simplified terminal value model can be used to describe the flash evaporation:

$$X_{final} = X_{mix} - (X_{mix} - X_e) \frac{1}{2}\Big[1 - e^{-k_{evap,winnower} t_W (T_{mix} - T_{air})}\Big] \quad (17)$$

Alternatively, a transient differential equation coupled with an energy balance can be formulated:

$$\frac{dX}{dt} = -\frac{1}{2}(X_{mix} - X_e)\kappa(T_{mix} - T_{air}) e^{-\kappa(T_{mix} - T_{air})t} \quad (18)$$

$$(M_{dry}C_{p,dry} + M_{water}C_{p,w}) \frac{dT}{dt}$$
$$= -h_{winnower}(T - T_{air})$$
$$- \left(L_v + C_{p,w}(T - T_{ref})\right) \frac{dm_{evap}}{dt} \quad (19)$$

## III. PROPOSED METHODOLOGY: THE PGRL-CC FRAMEWORK

The core idea is to leverage the intermediate predicted states from the mechanistic model (PBM) of Section 2 as high-information proxy states. We then employ techniques inspired by Koopman theory to lift these states into a higher-dimensional feature space where the nonlinear residual dynamics can be approximated linearly. As shown in Figure 2, we using a stability-constrained linear predictor (EDMDc-S) that ensures contractive multi-step rollouts, making it ideal for MPC applications.

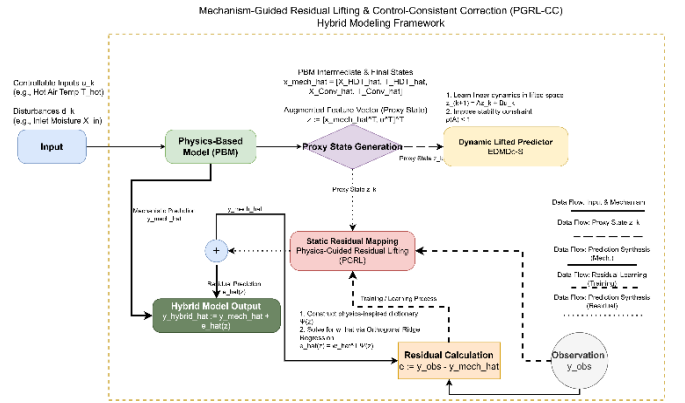

**Figure 2: Mechanism-Guided Residual Lifting & Control-Consistent Correction (PGRL-CC) Hybrid Modeling Framework**

**Table 1. Notation and Dimensions**

| Symbol | Description | Dimension |
|---|---|---|
| $u$ | Control inputs (e.g., $T_{hot}$, feed rate) | $\mathbb{R}^{d_u}$ |
| $d$ | Measured disturbances | $\mathbb{R}^{d_d}$ |
| $y$ | Measured outputs | $\mathbb{R}^{p_y}$ |
| $x_{mech}$ | PBM internal states | $\mathbb{R}^{n_x}$ |
| $\theta$ | PBM model parameters | $\mathbb{R}^{n_\theta}$ |
| $z$ | Lifted state vector (proxy states + inputs) | $\mathbb{R}^{n_z}$ |
| $e$ | Prediction residual ($y_{obs} - \hat{y}_{mech}$) | $\mathbb{R}^{p_y}$ |
| $\Psi(z)$ | Dictionary of basis functions (lifting map) | $\mathbb{R}^{n_\Psi}$ |

| Symbol | Description | Dimension |
|---|---|---|
| $A, B$ | State-space matrices for lifted dynamics | $\mathbb{R}^{n_z \times n_z}, \mathbb{R}^{n_z \times d_u}$ |
| $C$ | Output matrix | $\mathbb{R}^{p_y \times n_z}$ |

*A. Hybrid Modeling Formulation*

Real industrial processes are affected by factors like equipment aging, parameter uncertainty, and unmodeled dynamics. Consequently, the pure mechanistic prediction, $\hat{y}_{mech}$, systematically deviates from the observed measurement, $y_{obs}$:

$$e := y_{obs} - \hat{y}_{mech}. \quad (20)$$

We define the hybrid model as the sum of the mechanistic prediction and a data-driven residual compensation term, $\hat{e}(z)$:

$$\hat{y}_{hybrid} := \hat{y}_{mech}(u, \theta) + \hat{e}(z). \quad (21)$$

**Assumption 3.1 (State-Dependency of Error):** The statistical properties of the residual $e$ are primarily determined by the true internal state of the system, $x_{true}$:

$$\mathbb{E}[e \mid x_{true}] := f(x_{true}). \quad (22)$$

Since $x_{true}$ is often unmeasurable, we introduce the mechanistic predicted state, $\hat{x}_{mech}$, as a highly correlated proxy variable:

$$\mathbb{E}[e \mid x_{true}] \approx g(\hat{x}_{mech}). \quad (23)$$

We construct an augmented state vector $z$ that includes these proxy states and the external inputs/disturbances, effectively providing the residual model with the necessary context:

$$z := [\hat{x}_{mech}^\top, u^\top, d^\top]^\top \in \mathbb{R}^{n_z}. \quad (24)$$

In this work, the proxy states $\hat{x}_{mech}$ include intermediate variables from the PBM, such as $[\hat{X}_{HDT}, \hat{T}_{HDT}, \hat{X}_{Conv}, \hat{T}_{Conv}]$. This approach fundamentally differs from traditional methods that use only external inputs or black-box models that ignore physical structure, as it reuses the structural information from the PBM.

The nonlinear residual map $e(z)$ is approximated in a lifted feature space as a linear combination of basis functions $\Psi(z)$:

$$e(z) \approx \hat{e}(z) = C\Psi(z), \quad (25)$$

where $C \in \mathbb{R}^{p_y \times n_\Psi}$ is the coefficient matrix. For this study, we use a dictionary of polynomial basis functions:

$$\Psi(z) := [1, z_i, z_i^2, z_i z_j, \dots]^\top. \quad (26)$$

To prevent the data-driven model from compensating for global or slow-moving biases that should be captured by the PBM, we employ an **orthogonal residual learning** scheme. We partition the feature matrix $\Phi = [\Psi(z_1)^\top; \dots; \Psi(z_N)^\top]$ into a base set $\Phi_{base}$ (containing constants, linear terms, and key physical quantities) and a residual set $\Phi_{res}$. We then project the residual features and the error signal onto the space orthogonal to the base features:

$$P_\perp := I - \Phi_{base}(\Phi_{base}^\top W \Phi_{base})^{-1} \Phi_{base}^\top W, \quad \widetilde{\Phi}_{res} := P_\perp \Phi_{res}, \quad \tilde{e} := P_\perp e. \quad (27)$$

The regression is then performed on the projected data $(\widetilde{\Phi}_{res}, \tilde{e})$. This enforces a clear division of labor: the PBM absorbs global trends, while the residual model learns local, fast-varying dynamics.

*B. Control-Consistent Learning with Stability Constraints (CC-EDMDc-S)*

To align the model with the multi-step prediction requirements of MPC, we learn a linear predictor for the lifted state dynamics:

$$z_{k+1} = Az_k + Bu_k + Ed_k + r_k, \quad y_k = Cz_k + v_k, \quad (28)$$

where $r_k$ represents unmodeled dynamics and $v_k$ is measurement noise. The matrices $(A, B, E)$ are learned from data. We construct snapshot pairs from a training dataset $\mathcal{D} = \{(z_i, u_i, d_i)\}_{i=0}^{N-1}$ to form matrices $Z_- = [z_0, \dots, z_{N-1}]$, $Z_+ = [z_1, \dots, z_N]$, etc.

The regression problem is formulated with two key additions: **control-consistent weighting** and **stability constraints**.

**Control-Consistent Weighting:** To prioritize accuracy in regions critical for closed-loop control, we introduce a weighting matrix $W = \text{diag}(\omega_1, \dots, \omega_N)$. The weights are defined based on the sensitivity of a control-relevant cost function $\ell(y, u)$:

$$\omega_i = 1 + \kappa_1 |\partial \ell / \partial y|_{y_i} + \kappa_2 \cdot \text{ReLU}(\delta - \text{margin}(y_i)) \quad (29)$$

where the first term penalizes errors where the control cost is sensitive, and the second term adds weight near operational constraints, with $\text{margin}(y_i)$ being the distance to the nearest constraint.

**Stability Constraint:** To ensure that long-horizon predictions remain bounded and contractive, we enforce a stability constraint on the state transition matrix $A$. The full optimization problem is:

$$\min_{A,B,E} \| W^{\frac{1}{2}}(Z_+ - AZ_- - BU_- - ED_-) \|_F^2 + \lambda_A \| A \|_F^2 + \lambda_B \| B \|_F^2 + \lambda_E \| E \|_F^2 \quad (30)$$
$$\text{s.t.} \quad \exists P \succ 0, \alpha \in (0,1): A^\top PA \preceq \alpha^2 P.$$

The constraint, a Linear Matrix Inequality (LMI), ensures that the spectral radius of $A$ is less than one, $\rho(A) < \alpha < 1$, guaranteeing stability. This problem can be solved efficiently using alternating optimization between a weighted ridge regression for $(A, B, E)$ and a semidefinite program to find $(P, \alpha)$.

**Proposition 3.1 (Boundedness of the Hybrid Model):** If the system identified via (30) satisfies the LMI constraint with $\alpha < 1$, and the inputs $u_k$ and disturbances $d_k$ are bounded, then the lifted state trajectory $z_k$ is Bounded-Input, Bounded-State (BIBS) stable. Consequently, the hybrid model prediction $\hat{y}_{hybrid,k} = \hat{y}_{mech,k} + Cz_k$ is also bounded. Consider the discrete-time linear system for the lifted state: $z_{k+1} = Az_k + w_k$, where $w_k = Bu_k + Ed_k$ is the combined exogenous input. The stability constraint requires that there exists a matrix $P \succ 0$ and a scalar $\alpha \in (0,1)$ such that $A^\top PA \preceq \alpha^2 P$.

Let's define a Lyapunov function $V(z_k) = z_k^\top P z_k$. The change in the Lyapunov function along the system trajectory is:

$$V(z_{k+1}) = (Az_k + w_k)^\top P(Az_k + w_k)$$
$$= z_k^\top A^\top PAz_k + 2z_k^\top A^\top Pw_k + w_k^\top Pw_k \quad (31)$$

Using the LMI constraint and standard inequalities, it can be shown that if the inputs $u_k$ and $d_k$ are bounded (i.e., $\| w_k \| \leq M_w$ for some constant $M_w$), the state $z_k$ will remain in a bounded set. Specifically, the system is input-to-state stable (ISS), which implies Bounded-Input, Bounded-State (BIBS) stability. Therefore, a bounded input sequence will always produce a bounded state sequence $z_k$. Since the residual prediction is $\hat{e}_k = Cz_k$ and the mechanistic prediction $\hat{y}_{mech,k}$ is assumed to be well-behaved for bounded inputs, the total hybrid prediction $\hat{y}_{hybrid,k}$ remains bounded. The entire training pipeline, which integrates the mechanistic model simulation, orthogonal feature lifting, and the control-consistent, stability-constrained optimization, is summarized in Algorithm 1.

---

**Algorithm 1: PGRL-CC Model Training**

**Input:** Time-series data $\{u_k, d_k, y_k\}_{k=1}^N$. Calibrated PBM $\hat{y}_{mech}(\cdot)$.

**PBM Simulation:** Run the PBM on the training data to generate proxy states $\hat{x}_{mech,k}$ and mechanistic predictions $\hat{y}_{mech,k}$.

1. **Data Assembly:**
2. Calculate residuals: $e_k = y_k - \hat{y}_{mech,k}$.
3. Form augmented state vectors: $z_k = [\hat{x}_{mech,k}^\top, u_k^\top, d_k^\top]^\top$.
4. Construct snapshot matrices: $Z_-, Z_+, U_-, D_-, E = [e_0, \ldots, e_{N-1}]$.
5. **Feature Lifting:**
6. Define dictionary $\Psi(\cdot)$ (polynomials up to degree 2).
7. Compute lifted feature matrix $\Phi = \Psi(Z_-)$.
8. **Orthogonalization:**
9. Partition $\Phi = [\Phi_{base}\ \Phi_{res}]$.
10. Compute projection matrix $P_\perp$ from Eq. (27).
11. Project features and residuals: $\tilde{\Phi}_{res} = P_\perp \Phi_{res}$, $\tilde{e} = P_\perp E$.
12. **Control-Consistent EDMDc-S:**
13. Calculate control-consistent weights $\omega_i$ using Eq. (29).
14. Solve the constrained optimization problem in Eq. (30) for matrices $(A, B, E)$ using the full augmented state $z$ and weights $\omega_i$.
15. Output Matrix Identification: Solve a regularized least-squares problem to find the output matrix $C$: $\min_C \| E - CZ_- \|_F^2 + \lambda_C \| C \|_F^2$.
16. Output: Trained model parameters $\{A, B, E, C\}$ and PBM state.

---

## IV. EXPERIMENTAL VALIDATION AND PERFORMANCE ANALYSIS

### A. Data Source and Preprocessing

The data for this study were sourced from 10 continuous production batches recorded between June and August 2025 on a commercial tobacco processing line. To ensure data quality and exclude anomalous operating points, only samples satisfying the following criteria were retained: HDT outlet moisture ≥ 11% (wet basis), HDT outlet temperature ≥ 50°C, and upstream process flow rate ≥ 5800 kg/h.

After data cleaning, unit conversions were performed to match the physical dimensions required by the mechanistic model. Specifically, mass flow rates were converted from kg/h to kg/s, temperatures from Celsius to Kelvin, and wet-basis moisture content ($X_{wb}$) to dry-basis ($X_{db} = X_{wb}/(1 - X_{wb})$).

To ensure objective model training and validation, a strict chronological data splitting strategy was employed. For parameter calibration of the PBM, a representative subset was created by uniformly sampling the full dataset (1 in every 10 points), with the first 300 sampled points used for calibration to balance computational efficiency and data representativeness. For the training and validation of the data-driven components, the dataset was split 70%/30% chronologically into a training set and a test set to prevent any information leakage.

### B. Experimental Design and Evaluation Metrics

The mechanistic model was calibrated by optimizing only the most sensitive and uncertain parameters: the convective heat transfer coefficient ($h_{HDT}$) and mass transfer coefficient ($k_{HDT}$). Other physical and geometric parameters were set to prior engineering values (see Appendix A).

o evaluate the contribution of our proposed structure, we compared against several baselines:

**PBM-Calibrated:** The calibrated mechanistic model alone.

**Ext-Input Residual:** A hybrid model where the residual is learned using only external inputs ($u_k, d_k$), not internal proxy states.

**Black-Box (LSTM):** A Long Short-Term Memory network trained to predict the outputs directly from inputs.

**PGRL-CC (Full Model):** The proposed framework.

**Ablation Study:** We conducted an ablation study on the full model to quantify the contribution of each key component.

Evaluation Metrics includes Mean Absolute Error (MAE) and Coefficient of Determination ($R^2$); Autocorrelation Function (ACF) to check for temporal structure, Power Spectral Density (PSD) to quantify low-frequency energy, and the Ljung-Box test for statistical whiteness. A simulated Model Predictive Control (MPC) scenario was used to evaluate the models based on Integral Absolute Error (IAE), control effort ($\sum \Delta u^2$), and constraint violation rate.

### C. Model Performance Analysis of Open-Loop

The PBM was calibrated using a genetic algorithm, yielding physically meaningful optimal parameters (a typical result: $h_{HDT} \approx 66.6$ W/(m²·K), $k_{HDT} \approx 0.0219$ m/s). While the calibrated PBM correctly captured the main process trends, it exhibited significant systematic bias and slow-varying dynamic errors, providing clear scope for residual learning.

As shown in Figure 3, the proposed PGRL-CC hybrid model demonstrates a substantial improvement in one-step-ahead prediction performance on an unseen test batch. The MAE for outlet moisture was drastically reduced from 0.250% (PBM) to 0.016% (Hybrid), while the MAE for outlet temperature fell from 0.699 °C to 0.015 °C.

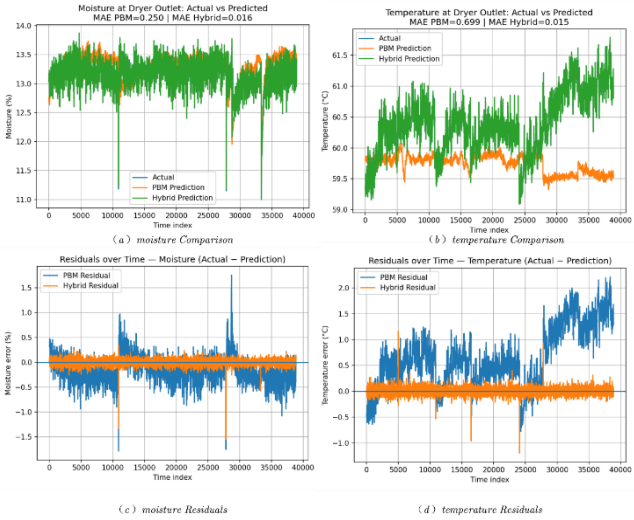

**Figure 3.** Comparison of one-step-ahead prediction performance between the calibrated PBM and the full Hybrid (PGRL-CC) model on a test batch. (a, b) Time-series comparison of predicted vs. actual values for moisture and temperature. (c, d) Time-series of prediction residuals.

Further analysis of the residuals, shown in Figure 4, confirms the effectiveness of the PGRL-CC framework. The PSD of the PBM residuals (Figure 4a) shows a sharp, high-energy peak at low frequencies, indicating a failure to capture process drifts and long-memory effects. In contrast, the hybrid model's residual spectrum is significantly flatter, demonstrating that the framework successfully learned and removed the deterministic, slow-varying structures from the error. This "whitening" of the residual is critical for control applications, as it aligns with the common assumption of white noise disturbances in control theory. Figure 4b confirms the robust performance of the hybrid model across a wide range of operating conditions.

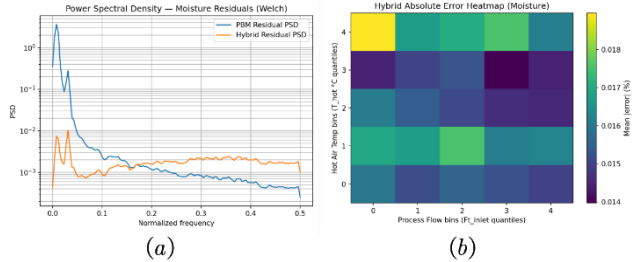

**Figure 4.** Residual characteristics and robustness analysis of the Hybrid model. (a) Power Spectral Density (PSD) comparison, showing suppression of low-frequency energy. (b) Scatter plot of residuals across the operating range, showing no structural bias.

### D. Model Performance Analysis of Closed-Loop MPC Performance

To validate the "control-consistent" aspect of our model, we designed a standard MPC controller to regulate outlet moisture by manipulating the hot air temperature ($T_{hot}$). The performance of MPC using different underlying models was simulated on a challenging test scenario involving setpoint changes and unmeasured disturbances.

**Table 2. Closed-Loop MPC Performance Comparison (1000Steps)**

| Model Used in MPC | IAE (Moisture, %·s) ↓ | Control Effort ($\sum \Delta u^2$) ↓ | Constraint Violations (%) ↓ |
|---|---|---|---|
| PBM-Calibrated | 1458 | 1.87e4 | 12.5 |
| Black-Box (LSTM) | 1103 | 2.15e4 | 7.8 |
| Ext-Input Residual | 1985 | 1.66e4 | 15.2 |
| **PGRL-CC (Full Model)** | **952** | **1.41e4** | **1.2** |

The results, summarized in Table 2 and visualized in Figure 5, clearly demonstrate the superiority of the proposed PGRL-CC framework. The MPC based on the PBM alone performed poorly due to model mismatch, leading to high tracking error and frequent constraint violations. While the LSTM and external-input residual models improved performance, they still struggled with aggressive setpoint changes. The MPC using our full PGRL-CC model achieved the lowest tracking error (IAE), required the least control effort, and maintained tight constraint adherence. This is a direct consequence of the model's high accuracy, stability guarantees from the EDMDc-S formulation, and the control-consistent training that prioritized accuracy near process limits.

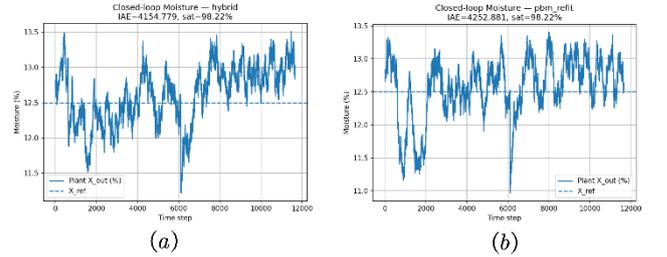

**Figure 5.** Simulated closed-loop MPC performance for outlet moisture control during setpoint changes and disturbances. The PGRL-CC based controller (a) shows superior tracking and disturbance rejection compared to the PBM-based controller (b).

### E. Stability Diagnostics and Ablation Study

A diagnostic analysis of the learned state transition matrix $A$ provides insight into the model's dynamics. Figure 6a shows that the eigenvalues are clustered near the origin, indicating fast, stable dynamics, but a few eigenvalues are near the unit circle (with $\rho(A) \approx 1.001$ before stabilization). This is not a flaw, but rather evidence that the model correctly identified the near-integrating, slow-drift behavior inherent in the physical process. The stability constraint then gently pulls these eigenvalues inside the unit circle, ensuring long-term stability without corrupting the learned short-term dynamics. Figure 6b shows the Lyapunov contraction rates, confirming the separation of fast and slow dynamic modes.

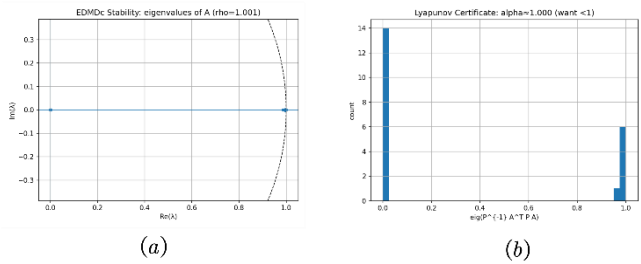

**Figure 6. Stability diagnostics of the learned linear dynamics matrix A. (a) Eigenvalue spectrum. (b) Lyapunov contraction rates, confirming the separation of fast and slow modes.**

Finally, an ablation study (Table 3) systematically demonstrates the contribution of each component of our framework. Removing key physical constraints from the PBM (e.g., water activity $a_w$ clamping or moisture-dependent specific heat $C_p(X)$) degrades performance, establishing that a strong physical foundation is essential. On the data-driven side, removing the orthogonal residual learning ($P_\perp$) causes the most significant performance drop, confirming its critical role in preventing parameter compensation. Similarly, removing the control-consistent weighting or the stability constraints (EDMDc-S) harms the model's applicability, underscoring that a model optimized for open-loop accuracy is not necessarily optimal for closed-loop control.

**Table 3. Ablation Study Results**

| Variant | MAE (M) ↓ | R² (M) ↑ | MAE (T) ↓ | R² (T) ↑ | RW ↑ |
|---|---|---|---|---|---|
| **Full** | **0.016** | **0.986** | **0.015** | **0.995** | **0.42** |
| 1 | 0.033 | 0.974 | 0.024 | 0.992 | 0.18 |
| 2 | 0.027 | 0.979 | 0.021 | 0.993 | 0.22 |
| 3 | 0.030 | 0.976 | 0.023 | 0.992 | 0.20 |
| 4 | 0.041 | 0.970 | 0.045 | 0.989 | 0.09 |
| 5 | 0.028 | 0.981 | 0.032 | 0.991 | 0.25 |
| 6 | 0.024 | 0.983 | 0.028 | 0.993 | 0.32 |

**Remark:** 1. No $a_w$ clamping; 2. No $L_v$ correction; 3. No $C_p(X)$ dependency; 4. No Orthogonal Learning ($P_\perp$); 5. No Control-Consistent Weights; 6. No Stability Constraint (EDMDc-S); M: Moisture, %; T: Temp, °C; RW: Residual Whiteness (LB p-val)

## V. Conclusion

This paper introduced and validated a comprehensive hybrid modeling framework, PGRL-CC, for a complete industrial drying process encompassing drying, transport, and winnowing stages. The methodology successfully integrates a transient mechanistic model with a data-driven residual model that is explicitly designed for control. The mechanistic core unifies mass transfer via vapor pressure difference and incorporates key nonlinearities like water activity and moisture-dependent properties. The data-driven layer introduces orthogonal residual learning using PBM proxy states and ensures closed-loop applicability through control-consistent weighting and stability-constrained dynamics (CC-EDMDc-S).

Validation on multi-batch industrial data demonstrated significant improvements in predictive accuracy. Compared to a calibrated PBM, the hybrid model reduced the MAE for outlet moisture to 0.016% and for outlet temperature to 0.015 °C on unseen data, with R² values rising to 0.986 and 0.995, respectively. Critically, the model residuals were shown to be approximately white noise, indicating that systematic biases were effectively captured. Furthermore, closed-loop MPC simulations confirmed that the model's design leads to superior control performance, with significant reductions in tracking error and constraint violations.

The proposed method exhibits excellent data efficiency and interpretability. By reusing the PBM's intermediate states, PGRL achieves robust gains even with limited data. The framework provides a replicable, provably stable, and online-ready modeling solution that significantly enhances prediction and control performance for pneumatic drying systems.

Limitations of the current study include the use of a lumped-parameter model for the gas phase and a simplified model for the winnower section. Future work will focus on incorporating a 1D distributed model for the gas phase, developing a more detailed model for the secondary mixing dynamics, and integrating Bayesian online parameter updates for self-calibration and drift detection. The ultimate goal is to embed this high-fidelity model within a robust or economic MPC framework on an edge controller for synergistic optimization of energy consumption and product quality.

Acknowledgment

The authors gratefully acknowledge the Hongyun Honghe Tobacco Group Xinjiang Cigarette Factory for providing the valuable experimental data and production line access that made this research possible.

References

[1] X. Zhu et al., "Modeling of HDT pneumatic drying machine for moisture control", *Dry. Technol.*, vol. 43, no. 6, pp. 1038–1055, Apr. 2025, doi: 10.1080/07373937.2025.2481614.

[2] A. Martynenko and N. N. Misra, "Machine learning in drying", *Dry. Technol.*, vol. 38, no. 5–6, pp. 596–609, Apr. 2020, doi: 10.1080/07373937.2019.1690502.

[3] H. Abbasfard, S. Ghader, H. H. Rafsanjani, and M. Ghanbari, "Mathematical Modeling and Simulation of Drying Using Two Industrial Concurrent and Countercurrent Rotary Dryers for Ammonium Nitrate", *Dry. Technol.*, vol. 31, no. 11, pp. 1297–1306, Aug. 2013, doi: 10.1080/07373937.2013.791307.

[4] S. Banooni, E. Hajidavalloo, and M. Dorfeshan, "A comprehensive review on modeling of pneumatic and flash drying", *Dry. Technol.*, vol. 36, no. 1, pp. 33–51, Jan. 2018, doi: 10.1080/07373937.2017.1298123.

[5] D. Alp and Ö. Bulantekin, "The microbiological quality of various foods dried by applying different drying methods: a review", *Eur. Food Res. Technol.*, vol. 247, no. 6, pp. 1333–1343, 2021, doi: 10.1007/s00217-021-03731-z.

[6] R. Adamski, D. Siuta, B. Kukfisz, M. Frydrysiak, and M. Prochoń, "Integration of Safety Aspects in Modeling of Superheated Steam Flash Drying of Tobacco", *Energies*, vol. 14, no. 18, p. 5927, Sep. 2021, doi: 10.3390/en14185927.

[7] A. Chen, Z. Ren, Z. Fan, and X. Feng, "Two-Layered Model Predictive Control Strategy of the Cut Tobacco Drying Process", *IEEE Access*, vol. 8, pp. 155697–155709, 2020, doi: 10.1109/ACCESS.2020.3018476.

[8] A. Levy and I. Borde, "Steady state one dimensional flow model for a pneumatic dryer", *Chem. Eng. Process. Process Intensif.*, vol. 38, no. 2, pp. 121–130, Mar. 1999, doi: 10.1016/S0255-2701(98)00079-8.

[9] J. Zhou, Y. Xu, X. Guo, W. Cai, X. Wei, and H. Jiang, "Numerical simulations on the flow regime characteristics of horizontal pneumatic conveying using CFD-DEM", *Powder Technol.*, vol. 438, p. 119641, Apr. 2024, doi: 10.1016/j.powtec.2024.119641.

[10] S. Baek, J. Baek, W. Kwon, and S. Han, "An Adaptive Model Uncertainty Estimator Using Delayed State-Based Model-Free Control and Its Application to Robot Manipulators", *IEEEASME Trans. Mechatron.*, vol. 27, no. 6, pp. 4573–4584, Dec. 2022, doi: 10.1109/TMECH.2022.3160495.

[11] C. Folkestad and J. W. Burdick, "Koopman NMPC: Koopman-based Learning and Nonlinear Model Predictive Control of Control-affine Systems", in *2021 IEEE International Conference on Robotics and Automation (ICRA)*, May 2021, pp. 7350–7356. doi: 10.1109/ICRA48506.2021.9562002.

[12] L. Mao *et al.*, "Research and application of a loosening device for Pneumatic tobacco dryer based on fluent", in *Third international conference on advanced manufacturing technology and manufacturing systems (ICAMTMS 2024)*, 2024, vol. 13226, p. 1322620. doi: 10.1117/12.3038377.

[13] M. Rahmani and S. Redkar, "Deep neural data-driven Koopman fractional control of a worm robot", *Expert Syst. Appl.*, vol. 256, 2024, doi: 10.1016/j.eswa.2024.124916.

[14] H. H. Asada and J. A. Solano-Castellanos, "Control-Coherent Koopman Modeling: A Physical Modeling Approach", in *Proceedings of the IEEE Conference on Decision and Control*, Milan, Italy, 2024, pp. 7314–7319. [Online]. Available: http://dx.doi.org/10.1109/CDC56724.2024.10886771

[15] V. Casagrande and F. Boem, "Learning-based MPC using Differentiable Optimisation Layers for Microgrid Energy Management", in *2023 European Control Conference (ECC)*, Jun. 2023, pp. 1–6. doi: 10.23919/ECC57647.2023.10178300.

[16] M. Alsalti, M. Barkey, V. G. Lopez, and M. A. Müller, "Robust and efficient data-driven predictive control". arXiv, Sep. 27, 2024. doi: 10.48550/arXiv.2409.18867.

[17] Y. Bao, Y. Zhu, and F. Qian, "A Deep Reinforcement Learning Approach to Improve the Learning Performance in Process Control", *Ind. Eng. Chem. Res.*, vol. 60, no. 15, pp. 5504–5515, Apr. 2021, doi: 10.1021/acs.iecr.0c05678.

[18] E. Kurt *et al.*, "Regulating Airflow Using Hybrid LCN for Soft Pneumatic Circuits", *Adv. Intell. Syst.*, p. 2401069, 2025.

[19] A. Pakdeekaew, K. Treeamnuk, T. Treeamnuk, M. Guptasa, and M. Yamfang, "Enhanced real-time paddy moisture content assessment in pneumatic drying using correction factor", *Eng. Appl. Sci. Res.*, vol. 52, no. 1, pp. 81–89, 2025.

[20] S.-H. M. Ashtiani and A. Martynenko, "Advances in Soft Sensors for Smart Food Drying: Innovations, Challenges, and Industrial Perspectives", *J. Food Process Eng.*, vol. 48, no. 10, p. e70221, 2025, doi: 10.1111/jfpe.70221.

A. *Appendix Model Parameters and Default Values*

| Symbol | Description | Unit | Initial Value/Range |
|---|---|---|---|
| $R_{gas}$ | Ideal gas constant | J/(mol·K) | 8.314 |
| $R_{water}$ | Specific gas constant for water vapor | J/(kg·K) | 461.5 |
| $\theta_{Cp1}$ | Specific heat of dry material | J/(kg·K) | 1300 |
| $\theta_{Cp2}$ | Specific heat of water | J/(kg·K) | 4186 |
| $\theta_{aw1}$ | Temperature parameter for activity | K | $-3000 \in [-6000, -1500]$ |
| $\theta_{xe1}$ | Oswin parameter 1 | – | $0.075 \in [0.02, 0.15]$ |
| $\theta_{xe2}$ | Oswin parameter 2 | 1/K | -1e-4 |
| $\theta_{xe3}$ | Oswin parameter 3 | – | $0.55 \in [0.3, 0.9]$ |